\documentclass[aip,jcp,twocolumn,reprint]{revtex4-1}

\usepackage{amssymb,amsmath}

\usepackage{graphicx}

\newcommand\beq{\begin{equation}}
\newcommand\eeq{\end{equation}}
\newcommand\beqa{\begin{eqnarray}}
\newcommand\eeqa{\end{eqnarray}}
\newcommand{\dd}{\text{d}}

\usepackage{color}

\begin{document}

\title{Homogeneous states in driven granular mixtures: Enskog kinetic theory versus molecular dynamics simulations}
\author{Nagi Khalil}
\email{nagi@us.es}
\affiliation{Departamento de
F\'{\i}sica, Universidad de Extremadura, E-06071 Badajoz, Spain}
\author{Vicente Garz\'o}
\email{vicenteg@unex.es}
\homepage{http://www.unex.es/eweb/fisteor/vicente/}
\affiliation{Departamento de
F\'{\i}sica, Universidad de Extremadura, E-06071 Badajoz, Spain}

\begin{abstract}
The homogeneous state of a binary mixture of smooth inelastic hard disks or spheres is analyzed. The mixture is driven by a thermostat composed by two terms: a stochastic force and a drag force proportional to the particle velocity. The combined action of both forces attempts to model the interaction of the mixture with a bath or surrounding fluid. The problem is studied by means of two independent and complementary routes. First, the Enskog kinetic equation with a Fokker-Planck term describing interactions of particles with thermostat is derived. Then, a scaling solution to the Enskog kinetic equation is proposed where the dependence of the scaled distributions $\varphi_i$ of each species on the granular temperature occurs not only through the dimensionless velocity $\mathbf{c}=\mathbf{v}/v_0$ ($v_0$ being the thermal velocity) but also through the dimensionless driving force parameters. Approximate forms for $\varphi_i$ are constructed by considering the leading order in a Sonine polynomial expansion. The ratio of kinetic temperatures $T_1/T_2$ and the fourth-degree velocity moments $\lambda_1$ and $\lambda_2$ (which measure non-Gaussian properties of $\varphi_1$ and $\varphi_2$, respectively) are explicitly determined as a function of the mass ratio, size ratio, composition, density and coefficients of restitution. Secondly, to assess the reliability of the theoretical results, molecular dynamics simulations of a binary granular mixture of spheres are performed for two values of the coefficient of restitution ($\alpha=0.9$ and 0.8) and three different solid volume fractions ($\phi=0.00785$, 0.1 and 0.2). Comparison between kinetic theory and computer simulations for the temperature ratio shows excellent agreement, even for moderate densities and strong dissipation. In the case of the cumulants $\lambda_1$ and $\lambda_2$, good agreement is found for the lower densities although significant discrepancies between theory and simulation are observed with increasing density.
\end{abstract}


\date{\today}
\maketitle

\section{Introduction}
\label{sec1}

It is well established that in rapid flow conditions the dissipative nature of granular matter is captured by a simple fluid of smooth, inelastic hard spheres. When the system is isolated and homogenized, it rapidly reaches a homogeneous cooling state (HCS) for which all the time dependence of the distribution function only occurs through the granular temperature. The existence of the HCS for a granular mixture was demonstrated years ago \cite{GD99} from the Enskog kinetic theory where it was shown that the existence of the above state necessarily requires that the cooling rates for the kinetic temperatures $T_i$ of each species must be the same. This yields a violation of the equipartition theorem since the partial temperatures of each species are different for mechanically different particles. The dependence of $T_i$ on the parameters of the system was obtained from an approximate solution of the Enskog  equation \cite{GD99} and the accuracy of this theoretical result was confirmed by Monte Carlo simulations of the Enskog equation \cite{MG02} as well as by molecular dynamics (MD) simulations of a binary mixture of inelastic hard spheres. \cite{DHGD02}

However, the HCS is a quite idealized situation since in general one has to feed energy into the system to keep it under rapid flow conditions. When the injected energy compensates for the energy lost by collisions, a non-equilibrium \emph{steady} state is achieved. This external energy can be supplied to the system from the boundaries (for example, from vibrating walls \cite{YHCMW02}),  by bulk driving (as in air-fluidized beds \cite{AD06,SGS05}) or by the presence of the interstitial fluid. \cite{MLNJ01,YSHLC03,WZXS08} In the former case, this way of supplying energy can be incorporated in a theoretical description by means of boundary conditions. The price to be paid when the granular fluid is \emph{locally} driven is that strong spatial gradients appear usually in the bulk domain and hence, the theoretical description of these situations turns out to be difficult.

On the other hand, under certain experimental conditions the bulk driving is homogeneous and its effect on grains can be modeled by the action of an external driving force that heats the system \emph{homogeneously}. This type of external forces are called ``thermostats'' \cite{EM90} and are very useful not only in computer simulations \cite{puglisi,zippelius}  but also to understand some experimental results. \cite{GSVP11,PGGSV12} In this paper, the granular fluid is driven by the action of a thermostat composed by two different terms: (i) a stochastic force with the form of a white noise with zero mean and finite variance where the particles are randomly kicked between collisions \cite{WM96} and (ii) a drag force proportional to the particle velocity. The stochastic force tries to mimic the energy transfer from the interstitial fluid to grains while the viscous drag force could model the friction of granular particles with the surrounding fluid. When the influence of granular particles  on the state of the surrounding fluid can be neglected, \cite{K90,KH01,GTSH12} the presence of thermostat leads to an additional operator in the Enskog equation, besides the Enskog collisional operator. In this paper, we identify the exact limit where the new operator adopts its commonly used Fokker-Planck form. This kind of thermostat, which has been widely used in previous works by other authors, \cite{puglisi} includes many of the thermostats commonly used in the literature of driven granular fluids. \cite{zippelius}

The theoretical analysis of the homogeneous steady state of a granular binary mixture driven by a stochastic bath with friction has been recently carried out \cite{NG13} in the context of the inelastic Enskog equation. For the sake of simplicity, non-Gaussian corrections to the homogeneous distribution functions  were neglected to evaluate the partial temperatures of each species. In addition, the set of transport coefficients of the driven mixture were also obtained in Ref.\ \onlinecite{NG13} by solving the kinetic equation by means of the Chapman-Enskog method \cite{CC70} for a dilute gas. In fact, the \emph{local} version of the inherently homogeneous state of our system emerges as the zeroth-order approximation of  the Chapman-Enskog expansion performed in Ref.\ \onlinecite{NG13}. Hence, one of the goals of the present work is to determine analytically the fourth-degree velocity moments (or fourth cumulants) $\lambda_i$ of the velocity distributions $f_i(\mathbf{v},t)$ $(i=1,2)$. The parameters $\lambda_i$ provide information on the deviation of the distributions of each species with respect to their Maxwellian forms. The evaluation of the fourth cumulants allows us to gauge the impact of non-Gaussian contributions to the distributions $f_i(\mathbf{v},t)$ on the temperature ratio.

Another interesting open problem is to assess the ability of the Enskog kinetic equation to describe homogeneous driven states in granular mixtures. In the present paper, we perform MD simulations for a binary mixture of inelastic hard spheres driven by a stochastic bath with friction. Three different values of density are considered, covering the dilute limit as well as moderate densities. Moreover, two values of the (common) coefficient of restitution are studied. From the simulations it is possible to compute the partial temperatures of each species and compare them with those obtained from an \emph{approximate} solution to the Enskog kinetic equation. The evaluation of the kinetic temperatures is likely the main objective of the paper. As an added value, we also compute the cumulants $\lambda_i$ of the distribution functions from MD simulations.  To the best of our knowledge, the only comparison between kinetic theory and MD simulations for the cumulants was carried out years ago in the limit case of a dilute monocomponent granular gas. \cite{Ricardo} Thus, this is the first time that the Enskog predictions for $\lambda_1$ and $\lambda_2$ are compared against MD for granular mixtures at moderate densities.

It must be remarked that the comparison carried out in this paper must be considered as a stringent test of the Enskog equation (with the inclusion of the Fokker-Planck term) since MD avoids any assumptions inherent in kinetic theory or approximations made in solving the corresponding kinetic equations. We show here that the dependence of the temperature ratio (which is related with the second-degree velocity moments of the distribution functions) on mechanical properties and state conditions exhibits an excellent agreement with predictions of the Enskog kinetic theory, including moderate densities and strong dissipation. On the other hand, in the case of the cumulants, the agreement is very good in the low-density regime although the discrepancies between theory and simulation increase with the density.

The plan of the paper is as follows. In Sec.\ \ref{sec2}, the kinetic equation describing a granular binary mixture driven by a stochastic bath with friction is derived. Section \ref{sec3} deals with the homogeneous state where a scaling solution is proposed that depends on granular temperature through two dimensionless parameters (dimensionless velocity and reduced noise strength). Analytic results for the partial temperatures and the cumulants are obtained from an approximate solution to the Enskog equation based on the truncation of a Sonine polynomial expansion. In Sec.\ \ref{sec4}, the Enskog predictions are compared with those obtained from MD simulations for different systems and coefficients of restitution. Finally, the paper ends in Sec.\ \ref{sec5} with a brief discussion on the relevance of the results reported here.

\section{Granular mixtures driven by a stochastic bath with friction}
\label{sec2}

We consider a granular multicomponent mixture of $N_i$ smooth hard spheres in $d$ dimensions with masses $m_i$ and diameters $\sigma_i$, where the subscript $i$ labels one of the $s$ mechanically different species. In general, collisions among all pairs are \emph{inelastic} and are characterized by independent constant normal coefficients of restitution $\alpha_{ij}=\alpha_{ji}$, where $\alpha_{ij}$ is the coefficient of restitution for collisions between particles of species $i$ and $j$, $0<\alpha_{ij}\leq 1$.  The elastic limit corresponds to $\alpha_{ij}=1$. In all of the following, attention is restricted to spatially homogeneous states. At a kinetic level, all the relevant information on the state of the system is given through the knowledge of the one-particle distribution function $f_i(\mathbf{v},t)$ of each species $(i=1,\ldots, s)$. The quantity $f_i(\mathbf{v},t) \dd\mathbf{v}$ gives the \emph{average} number of particles of species $i$ which at time $t$ are moving with velocities in the range $\dd\mathbf{v}$ about $\mathbf{v}$. In order to maintain a fluidized granular mixture, an external energy source is coupled to every particle of the mixture in the form of a thermal bath. Under these conditions and in the absence of gravity, the time evolution of the distributions $f_i$ obey the set of $s$-coupled kinetic equations
\begin{equation}
\partial_{t}f_i=\sum_{j=1}^s\; J_{ij}[\mathbf{v}|f_i,f_j]+ \mathcal{F}_i[\mathbf{v}|f_i], \quad i=1,\ldots, s.
\label{eq:1}
\end{equation}
As usual, the first term on the right hand side of Eq.\ \eqref{eq:1} refers to the change of $f_i$ due to collisions among the particles while the second term $\mathcal{F}_i$ accounts for the interaction of species $i$ with the external bath. Upon writing the new term $\mathcal{F}_i$, we are assuming that the action of bath on species $i$ depends only on its distribution $f_i$.

For moderate solid volume fractions, one can neglect the velocity correlations between particles which are about to collide (molecular chaos hypothesis), and $J_{ij}[\mathbf{v}|f_i,f_j]$ reduces to the Enskog collision operator \cite{GD99}
\begin{eqnarray}
\label{eq:2}
J_{ij}\left[\mathbf{v}|f_i, f_j\right] &=& g_{ij}\sigma_{ij}^{d-1}\int
\dd\mathbf{v}_{2}\int \dd\widehat{\boldsymbol {\sigma}}\Theta
(\widehat{\boldsymbol {\sigma}}\cdot \mathbf{v}_{12})(\widehat{
\boldsymbol {\sigma }}\cdot \mathbf{v}_{12})\nonumber\\
& & \times \left[ \alpha_{ij}^{-2}f_i(\mathbf{v}_{1}^{\prime
})f_j(\mathbf{v}_{2}^{\prime})-f_i(\mathbf{v}_{1})f_j(\mathbf{v}_{2})\right].
\end{eqnarray}
Here, $g_{ij}$ is the spatial pair correlation function for particles of species $i$ and $j$ at contact,  $\widehat{\boldsymbol {\sigma}}$ is a unit vector directed along the line of centers from the sphere of species $i$ to that of species $j$ at contact, $\Theta$ is the Heaviside step function, $\mathbf{v}_{12}=\mathbf{v}_1-\mathbf{v}_2$ is the relative velocity, and the restituting (``precollisional'') velocities $\mathbf{v}_{1}'$ and $\mathbf{v}_{2}'$ are related to the ``postcollisional'' velocities by
\begin{equation}
\begin{split}
& \mathbf{v}_{1}^{\prime }=\mathbf{v}_{1}-\mu _{ji}\left( 1+\alpha_{ij}^{-1}\right) (\widehat{{\boldsymbol {\sigma }}}\cdot \mathbf{g}_{12})\widehat{{\boldsymbol {\sigma }}},\\
& \mathbf{v}_{2}^{\prime }=\mathbf{v}_{2}+\mu _{ij}\left(1+\alpha_{ij}^{-1}\right) (\widehat{{\boldsymbol {\sigma }}}\cdot \mathbf{g}_{12})\widehat{ \boldsymbol {\sigma}},
\end{split}
\label{eq:3}
\end{equation}
where $\mu_{ij}= m_{i}/\left( m_{i}+m_{j}\right) $. Except for the presence of the factor $g_{ij}$ (which accounts for the increase of the collision frequency for collisions $i$-$j$ due to excluded volume effects), the Enskog operator \eqref{eq:2} is identical to the Boltzmann collision operator for a low-density mixture. For this reason, henceforth we will call $J_{ij}$ as the Enksog-Boltzmann collision operator.

The interaction between particles of species $i$ with the thermal bath is modeled by the term \cite{ka81,GMT12}
\begin{eqnarray}
  \label{eq:4}
  \mathcal{F}_i[\mathbf{v}|f_i]= \int \dd\mathbf{v}' & & \left[  W_{i,\Delta t}(\mathbf{v} | \mathbf{v}',t) f_i(\mathbf{v}',t)\right. \nonumber\\
  & & \left.-
  W_{i,\Delta t}(\mathbf{v}'| \mathbf{v},t) f_i(\mathbf{v},t) \right],
\end{eqnarray}
where $W_{i,\Delta t}(\mathbf{v}'|\mathbf{v},t)$ is a transition probability, or density probability per unit time that a particle of species $i$ with velocity $\textbf{v}$ at time $t$ collides with the bath during a time interval $\Delta t$ and changes its velocity to $\textbf{v}'$.  Note that $W_{i,\Delta t}(\mathbf{v}'|\mathbf{v},t)$ does not depend on the state of grains, and therefore our model is essentially different from other approaches where the interaction between bath and grains is only driven by the state of the latter. \cite{brito,brey} Furthermore, we assume that the transition probability $W_{i,\Delta t}$ changes the velocity of the particle of species $i$ following the rule
\beq
\label{v1}
 \mathbf{v}= \mathbf{v}^\text{det}+\mathbf{v}^\text{st},
 \eeq
where the deterministic contribution is
\begin{equation}
  \label{eq:5}
  \mathbf{v}^\text{det}=\mathbf{v} (1-\epsilon_i), \quad \epsilon_i=\exp\left(-\frac{\gamma_\text{b}}{m_i^{\beta}} \Delta t \right) -1,
\end{equation}
while the stochastic contribution is
\beq
\label{v2}
\mathbf{v}^\text{st}=\left(\frac{\xi_\text{b}^2}{m_i^\lambda} \Delta t\right)^{1/2} \mathbf{w},
\eeq
with $\mathbf{w}$ being a random vector of zero mean and unit variance. The model parameters $\gamma_\text{b}, \ \xi_\text{b}^2, \ \beta$, and $\lambda$ are assumed to be constants that depend on the physical situation considered.

The transition probabilities $W_{i,\Delta t}(\mathbf{v}'|\mathbf{v},t)$ corresponding to the rule \eqref{v1} can be written as
\beq
\label{v3}
W_{i,\Delta t}(\mathbf{v}'|\mathbf{v},t)=W_{i,\Delta t}(\mathbf{v}|\mathbf{v}',t)=\frac{1}{\Delta t} \left(\frac{\xi_\text{b}^2}{m_i^\lambda} \Delta t\right)^{-d/2}P(\textbf{w}),
\eeq
where $P(\textbf{w})$ is the distribution of the random variable $\mathbf{w}$.  Since
\beq
\label{v3bis}
f_i(\mathbf{v}',t)\; \dd\mathbf{v}'=(1-\epsilon_i)^{-d}f_i\left(\frac{\mathbf{v}-\mathbf{v}^\text{st}}{1-\epsilon_i},t\right)
\; \dd\mathbf{v},
\eeq
then Eq. \ \eqref{eq:4} can be rewritten as
\begin{eqnarray}
  \label{eq:7}
  \mathcal{F}_i[\mathbf{v}|f_i]= \frac{1}{\Delta t} \int \; \dd\mathbf{w} \ P(\mathbf{w})  & & \left[(1-\epsilon_i)^{-d}f_i\left(\frac{\mathbf{v}-\mathbf{v}^\text{st}}{1-\epsilon_i},t\right) \right. \nonumber\\
  & &
  \left.  -f_i(\mathbf{v},t) \right].
\end{eqnarray}
Equations \eqref{v1}--\eqref{v2} allows us to identify the typical time $\Delta t$ characterizing the
time collision between particles of species $i$ and bath, the typical time associated with the deterministic
part $\tau_i^{(\gamma)}=m_i^\beta/\gamma_\text{b}$, and the typical velocity $v_i^{(\xi)}= \left(\frac{\xi_\text{b}^2}{m_i^\lambda} \Delta t\right)^{1/2}$ associated with the stochastic part of the interaction between particles and bath. Assuming that $\Delta t$ is small enough to ensure that $\Delta t \ll \tau_i^{(\gamma)}$ and $v_i^{(\xi)}\ll \sqrt{2T/m_i}$, then one can use the approximation
\begin{eqnarray}
  \label{eq:9}
   & &  (1-\epsilon_i)^{-d}f_i\left(\frac{\mathbf{v}-\mathbf{v}^\text{st}}{1-\epsilon_i},t\right) -f_i(\mathbf{v},t) \simeq
    \mathbf{v}^\text{st} \cdot \frac{\partial f_i}{\partial \mathbf{v}}\nonumber\\
   & & +\frac{\gamma_\text{b}}{m_i^\beta}\frac{\partial}{\partial \mathbf{v}}\cdot \left[\mathbf{v}f_i\right] \Delta t+\frac{1}{2}\mathbf{v}_i^\text{st} \mathbf{v}_i^\text{st} : \frac{\partial^2 f_i}{\partial \mathbf{v} \partial \mathbf{v}} + {\cal O}\left(\Delta t^{3/2}\right).\nonumber\\
\end{eqnarray}
Substitution of the relation \eqref{eq:9} into Eq. \eqref{eq:7} and taking into account Eq. \ \eqref{v2}, one gets
\begin{equation}
  \label{eq:10}
  \mathcal{F}_i[\mathbf{v}|f_i]= \frac{\gamma_\text{b}}{m_i^\beta}\frac{\partial}{\partial \mathbf{v}}\cdot \left[\mathbf{v}f_i(\mathbf{r},\mathbf{v},t)\right] +\frac{\xi_\text{b}^2}{2m_i^\lambda } \frac{\partial^2 f_i(\mathbf{r},\mathbf{v},t)}{\partial v^2},
\end{equation}
where use has been made of the fact that $\mathbf{w}$ has zero mean and unit variance. Equation \eqref{eq:10} represents the Fokker-Planck operator of a stochastic bath with friction, with $\gamma_\text{b}/m_i^\beta$ being the drift coefficient and $\xi_\text{b}^2/m_i^\lambda$ the diffusion one. \cite{ka81} Note that Eq.\ \eqref{eq:10} is independent of the particular form of the distribution $P(\mathbf{w})$.

The parameters $\beta$ and $\lambda$ appearing in Eq.\ \eqref{eq:10} can be considered as free parameters of the model. Thus, in the case $\gamma_\text{b}=0$ and $\lambda=0$ our thermostat reduces to the (pure) stochastic thermostat employed in previous works \cite{DHGD02,BT02} for granular binary mixtures. On the other hand, the choice $\beta=1$ and $\lambda=2$ yields the conventional Fokker-Planck model for ordinary (elastic) mixtures. \cite{puglisi,H03} In this context, the model defined by Eq.\ \eqref{eq:10} generalizes previous driven models used in the granular literature and only particular values of the bath parameters $\beta$ and $\lambda$ will be taken at the end of the calculations.

\section{Homogeneous states for granular binary mixtures}
\label{sec3}

We consider now a driven binary granular mixture ($s=2$). In the case that $\Delta t$ is small, as seen in Section \ref{sec2}, the distribution functions $f_1$ and $f_2$ verify the Enskog-Boltzmann kinetic equations
\begin{eqnarray}
 \label{eq:11}
\partial_{t}f_i(\mathbf{v},t)&-&\frac{\gamma_\text{b}}{m_i^\beta} \frac{\partial}{\partial\mathbf{v}}\cdot [\mathbf{v}
f_i(\mathbf{v},t)]-\frac{1}{2}\frac{\xi_\text{b}^2}{m_i^\lambda}\frac{\partial^2}{\partial v^2}f_i(\mathbf{v},t)\nonumber\\
&=&\sum_{j=1}^2\; J_{ij}[\mathbf{v}|f_i,f_j], \quad (i=1,2).
\end{eqnarray}

The partial densities $n_i$ and the granular temperature $T$ are defined, respectively, as
\begin{equation}
\label{eq:13}
n_i(t)=\int\; \dd\mathbf{v}\; f_{i}(\mathbf{v},t),
\end{equation}
\beq
\label{temp}
T(t)=\frac{1}{n}\sum_{i=1}^2\; \frac{m_i}{d}\int\; \dd \mathbf{v}\; v^2 f_{i}(\mathbf{v},t),
\eeq
where $n=n_1+n_2$ is the total number density. Apart from the global temperature $T$, the partial temperatures $T_i(t)$ associated with the kinetic energy of species $i$ are also properties of primary interest in \emph{granular} mixtures. They are defined as
\beq
\label{tempi}
T_i(t)=\frac{m_i}{d n_i}\int\; \dd \mathbf{v}\; v^2 f_{i}(\mathbf{v},t).
\eeq
The time evolution of $T(t)$ follows from the set of Enskog-Boltzmann equations \eqref{eq:11} that give \cite{NG13}
\beq
\label{balance}
\partial_t T=-2\gamma_\text{b}\sum_{i=1}^2
\frac{x_i T_i}{m_i^\beta}+\frac{\xi_\text{b}^2}{n}\sum_{i=1}^2\frac{\rho_i}{m_i^\lambda}-\zeta \,T,
\eeq
where $x_i=n_i/n$ is the mole fraction of species $i$, $\rho_i=m_in_i$ is the mass density of species $i$ and $\zeta$ is the total cooling rate due to inelastic collisions among all species. It is defined as
\beq
\label{zeta}
\zeta=T^{-1}\sum_{i=1}^2\;x_iT_i\zeta_i,
\end{equation}
where
\begin{equation}
\label{zetai} \zeta_i=-
\frac{m_i}{dn_iT_i}\sum_{j=1}^2\int \dd\mathbf{ v}\; v^{2}J_{ij}[{\bf
v}|f_{i},f_{j}]
\end{equation}
is the partial cooling rate for the partial temperature $T_i$. Analogously, the evolution equation for the partial temperatures $T_i$ can be obtained by multiplying both sides of Eq.\ \eqref{eq:11} by $m_iv^2/2$ and integrating over $\mathbf{v}$. The result is
\begin{equation}
\label{balancei}
\partial_tT_i=-\frac{2 T_i}{m_i^\beta}\gamma_\text{b} +\frac{\xi_\text{b}^2}{m_i^{\lambda-1}}-\zeta_i T_i.
\end{equation}

As noted in the dilute case, \cite{NG13} for given values of the driven parameters of the model (including $\beta$ and $\lambda$), the general solution to Eq.\ \eqref{eq:11} depends on velocity and time as well as the model parameters $\gamma_\text{b}$ and $\xi^2_\text{b}$. Thus, $f_i$ has the scaled form
\begin{equation}
\label{eq:12}
f_i(\mathbf{v}, \gamma_\text{b}, \xi_\text{b}^2, t)=n_iv_0(t)^{-d} \varphi_{i}\left(\mathbf{c}, \gamma^*, \xi^* \right),
\end{equation}
where $v_0(t)=\sqrt{2T(t)/\overline{m}}$ with $\overline{m}=m_1m_2/(m_1+m_2)$ and the reduced distribution $\varphi_{i}$ is an unknown function of the dimensionless parameters
\begin{equation}
\label{eq:15}
\mathbf{c}=\frac{\mathbf{v}}{v_0}, \quad
    \xi^*= \frac{\xi_{\text{b}}^2}{n\sigma_{12}^{d-1}\overline{m}^{\lambda-1} Tv_0},
\end{equation}
and
\beq
\label{gamma}
\gamma^*= \frac{\gamma_\text{b}}{n_{\text{s}}\sigma_{12}^{d-1}\overline{m}^{\beta}v_0}.
\eeq
The (reduced) drag parameter $\gamma^*$ can be easily written in terms of the (reduced) noise strength $\xi^*$ as
\beq
\label{omega}
\gamma^*=\omega^* \xi^{*1/3}, \quad
\omega^*= \frac{\gamma_\text{b}}{\overline{m}^\beta} \left(\frac{\overline{m}^\lambda}{2 \xi_{\text{b}}^2}\right)^{1/3} \left(n\sigma_{12}^{d-1}\right)^{-2/3}.
\eeq
We recall that while $\mathbf{c}$ and  $\xi^*$ are functions of time through its dependence on $T(t)$, $\omega^*$ is a constant parameter since the number density $n$ is also constant.

Substitution of the form \eqref{eq:12} into the Enskog-Boltzmann equation \eqref{eq:11} yields the following equation for the scaling distributions $\varphi_i(\mathbf{c},\omega^*, \xi^*)$:
\begin{eqnarray}
  \label{eq:16}
& & \Lambda^* \left[\frac{1}{2}\frac{\partial}{\partial\mathbf{c}}\cdot (\mathbf{ c}\varphi_i)+
\frac{3}{2}\xi^*\frac{\partial \varphi_i}{\partial \xi^*}\right]
-\frac{\omega^* \xi^{*1/3}}{M_i^\beta}\frac{\partial}{\partial \mathbf{ c}}\cdot \mathbf{ c} \varphi_i \nonumber\\
& &
- \frac{1}{4}\frac{\xi^*}{M_i^\lambda} \frac{\partial^2}{\partial c^2}\varphi_i=\sum_{j=1}^2
J_{ij}^*[\mathbf{c}|\varphi_i,\varphi_j],
\end{eqnarray}
where $M_i=m_i/\overline{m}$, $\Lambda^*=x_1\Lambda_1^*+x_2\Lambda_2^*$, and
\begin{equation}
  \label{eq:18}
  \Lambda^*_i = 2 \omega^* \xi^{*1/3} \frac{\chi_i}{M_i^\beta}- \frac{\xi^*}{M_i^{\lambda-1}} +\chi_i \zeta_i^*.
\end{equation}
Here, $\chi_i\equiv T_i/T$,
\beq
\label{zetaired}
\zeta_i^*\equiv \frac{\zeta_i}{n_i \sigma_{12}^{d-1}v_0}=-\frac{2}{d}\frac{M_i}{\chi_i}\sum_{j=1}^2\; \int\;
\dd\mathbf{c}\; c^2\; J_{ij}^*[\mathbf{c}|\varphi_i,\varphi_j],
\eeq
and the dimensionless Enskog-Boltzmann collision operator $J_{ij}^*[\mathbf{c}|\varphi_{i},\varphi_{j}]$ is given by
\begin{eqnarray}
  \label{eq:17}
  & & J_{ij}^*[\mathbf{c}|\varphi_{i},\varphi_{j}]= g_{ij} x_{j} \left(\frac{\sigma_{ij}}{\sigma_{12}}\right)^{d-1}\int \dd\mathbf{c}_{2}\int \dd\widehat{\boldsymbol {\sigma}}\Theta (\widehat{\boldsymbol {\sigma}}\cdot \mathbf{v}_{12}^*) \nonumber\\
  & & \times (\widehat{ \boldsymbol {\sigma }}\cdot \mathbf{v}_{12}^*) \left[ \alpha_{ij}^{-2}\varphi_{i}(\mathbf{c}_{1}^{\prime})\varphi_{j}(\mathbf{c}_{2}^{\prime})-\varphi_{i} (\mathbf{c}_{1})\varphi_{j}(\mathbf{c}_{2})\right],
\end{eqnarray}
where $\mathbf{v}^*_{12}=\mathbf{c}_1-\mathbf{c}_2$. It is worthwhile remarking that the functional dependence of the scaled distributions $\varphi_i$ on the variables $\mathbf{c}$, $\omega^*$ and $\xi^*$  is consistent with Eq. \eqref{eq:16}. As a consequence, the dependence of the temperature ratios $\chi_i$ on time is only through the dimensionless noise strength $\xi^*$. According to Eq.\ \eqref{balancei}, the evolution equation of $\chi_i$ is
\begin{equation}
  \label{eq:21}
\frac{3}{2} \Lambda^* \xi^* \frac{\partial \chi_i}{\partial \xi^*}=\chi_i \Lambda^*-\Lambda_i^*.
\end{equation}
Equation \eqref{eq:21} turns out to be a highly non-linear differential equation since the functions $\Lambda_i^*$ present an intricate  nonlinear dependence on $\chi_i$, even in the simplest Gaussian approximation to $\varphi_i$.

In summary, the solution to the homogeneous problem is defined by the two equations \eqref{eq:16} and the non-linear differential equation \eqref{eq:21}. These three equations must be solved self-consistently for the two scaled distributions $\varphi_1$ and $\varphi_2$ and the temperature ratio $\chi_1$ (since $\chi_2=(1-x_1 \chi_1)/x_2$). An approximate solution is described in the next subsection.

\subsection{Approximate solution}

A convenient way of characterizing $\varphi_i(\textbf{c}, \omega^*, \xi^*)$ in the range of low and intermediate velocities is through an expansion in a complete set of polynomials $\left\{P_q\right\}$ with a Gaussian measure. The coefficients $\lambda_q$ of such an expansion are polynomial moments of the distributions $\varphi_i$. In practice, the generalized Laguerre or Sonine polynomials \cite{AS72} are used. Approximate solutions for the moments $\lambda_q$ can be obtained by truncating the series at a given order. This approach is analogous to the moment method used for solving kinetic equations for ordinary gases. The idea has been also applied to inelastic systems for undriven and driven monocomponent gases   \cite{NE98,MS00,SM09,BP06,MVG13} and also in the case of free evolving multicomponent granular gases. \cite{GD99} In both cases, an excellent approximation has been to retain only the first two terms and the theoretical predictions compare quite well with Monte Carlo simulations. \cite{MG02,NE98,MS00,SM09,MVG13} A similar approximation is assumed here and hence, $\varphi_i$ is given by
\begin{equation}
\label{eq:23}
\varphi_i(\mathbf{c}) \to   \varphi_{i,\text{M}}(\mathbf{c}) \left\{1 +\frac{\lambda_i}{4}\left[ \theta_i^2 c^4-(d+2)\theta_i c^2+\frac{d(d+2)}{4}\right]\right\},
\end{equation}
where
\begin{equation}
\label{eq:24}
\varphi_{i,\text{M}}(\mathbf{c})=\pi^{-d/2} \theta_i^{d/2}\; e^{-\theta_i c^2}
\end{equation}
is the Maxwellian distribution with $\theta_i=M_i/\chi_i$. The dependence of $\varphi_{i,\text{M}}$ on the partial temperature $T_i$  is required by the definition \eqref{tempi}. The fourth cumulants $\lambda_i$ are defined as
\begin{equation}
  \label{eq:26} \lambda_{i}=2\left[\frac{4}{d(d+2)}\theta_i^2\langle c^4\rangle_i-1\right].
\end{equation}
with
\begin{equation}
\label{eq:27}
\langle c^k\rangle_i=\int\; \dd\mathbf{c}\; c^k \varphi_i.
\end{equation}
The coefficients $\lambda_i$ measure the deviation of $\varphi_i$ from their Maxwellian forms $\varphi_{i,\text{M}}$.

At this level of approximation, the unknown quantities are the temperature ratio $\chi_1$ and the cumulants $\lambda_1$ and $\lambda_2$. The equation determining $\chi_1$ is given by Eq. \eqref{eq:21} with $i=1$ while the cumulants can be obtained by multiplying the set of Boltzmann-Enskog equations \eqref{eq:16} by $c^4$ and integrating over velocity. The result is
\begin{eqnarray}
\label{eq:28}
& & \Lambda^*\left(1+\frac{1}{2}\lambda_i-\frac{3}{8}\xi^*\frac{d\lambda_i}{d\xi^*}\right)
-\left(\Lambda^*+\frac{\xi^*}{M_i^{\lambda-1}\chi_i}-\zeta_i^*\right)\nonumber\\
& & \times
\left(1+\frac{1}{2}\lambda_i\right)+\frac{\xi^*}{M_i^{\lambda-1}\chi_i}=
-\frac{2\theta_i^2}{d(d+2)}\Sigma_{i},
\end{eqnarray}
where
\begin{equation}
\label{eq:29}
\Sigma_i = \sum_{j=1}^2\; \int\; \dd\mathbf{c}\; c^4\; J_{ij}^*[\mathbf{c}|\varphi_i,\varphi_j],
\end{equation}
and use has been made of the results $\langle c^{2} \rangle_i=\frac{d}{2}\theta_i^{-1}$ and
\begin{equation}
\label{eq:29.1}
\int\; d{\bf c}\; c^{2p}\; \frac{\partial}{\partial {\bf
c}}\cdot {\bf c}\varphi(\mathbf{c})=-2p\langle c^{2p} \rangle_i,
\end{equation}
\begin{equation}
\label{eq:29.2}
\int\; \dd {\bf c}\; c^{2p}\; \frac{\partial^2}{\partial c^2}\varphi(\mathbf{c})=2p(2p+d-2)\langle c^{2p-2} \rangle_i.
\end{equation}

The set of coupled Eqs. \eqref{eq:21} and \eqref{eq:28} for $i=1,2$ are \emph{still} exact since we have not made use of the explicit form of the leading Sonine form \eqref{eq:23}. From experience with previous results, \cite{NE98,MS00,GD99} it is expected that the $\lambda_i$ are very small and hence, only linear terms in $\lambda_i$ are retained. Thus, approximate forms for the collision integrals defining $\zeta_i^*$ and $\Sigma_i$ can be obtained when one substitutes the first Sonine approximation \eqref{eq:23} into Eqs. \eqref{zetaired} and \eqref{eq:29} and neglects nonlinear terms in $\lambda_i$. The expressions of $\zeta_i^*$ and $\Sigma_i$ for an arbitrary number of dimensions are provided in Appendix \ref{appen1}. In compact form, they can be written as
\begin{equation}
\label{eq:31}
\zeta_1^*=\zeta_{10}+\zeta_{11}\lambda_1+\zeta_{12}\lambda_2,
\end{equation}
\begin{equation}
\label{eq:31.1}
\Sigma_1=\Sigma_{10}+\Sigma_{11}\lambda_1+\Sigma_{12}\lambda_2,
\end{equation}
where the quantities $\zeta_{ij}$ and $\Sigma_{ij}$ are given in Appendix \ref{appen1}. The forms of $\zeta_2^*$ and $\Sigma_2$ can be easily inferred from Eqs. \eqref{eq:31} and \eqref{eq:31.1}  by interchanging $1$ and $2$. Note that $\zeta_{ij}$ and $\Sigma_{ij}$ depend on $\xi^*$ through their dependence on the temperature ratio $\chi_1$.

The problem has been now reduced to quadratures and the solutions can be achieved as follows: (i) substitution of the relations \eqref{eq:31} into Eqs.\ \eqref{eq:21} and \eqref{eq:28} yields a system of nonlinear differential equations for $\chi_1$, $\lambda_1$, and $\lambda_2$ whose numerical integration provides the above quantities in terms of the (reduced) noise strength $\xi^*$; (ii) then the time dependence of the granular temperature $T(t)$ is obtained by numerically solving Eq.\ \eqref{balance}
and (iii) finally, all the quantities involved in the problem are obtained as a function of time since $\xi^*$ depends on $t$ through $T(t)$.

\subsection{Homogeneous steady states}

For arbitrary initial conditions, the simulations show that the system reaches after a transient regime a steady state. In this case, $\Lambda^*=\Lambda_1^*=\Lambda_2^*=0$ and the set of Eqs. \eqref{eq:28} become
\beqa
\label{lambda1}
& & \left(\frac{1}{2}\zeta_{10}+\frac{2\theta_1^2}{d(d+2)}\Sigma_{11}\right)
\lambda_1+\frac{2\theta_1^2}{d(d+2)}\Sigma_{12}\lambda_2 \nonumber\\
& & =
\frac{2\omega^*\xi^{*1/3}}{M_1^\beta}-\frac{2\theta_1^2}{d(d+2)}\Sigma_{10},
\eeqa
\beqa
\label{lambda2}
& & \frac{2\theta_2^2}{d(d+2)}\Sigma_{21}\lambda_1+\left(\frac{1}{2}\zeta_{20}+\frac{2\theta_2^2}
{d(d+2)}\Sigma_{22}\right)
\lambda_2 \nonumber\\
& &  =\frac{2\omega^*\xi^{*1/3}}{M_2^\beta}-\frac{2\theta_2^2}{d(d+2)}\Sigma_{20}.
\eeqa
Upon deriving Eqs.\ \eqref{lambda1} and \eqref{lambda2} use has been made of the expansions \eqref{eq:31} and \eqref{eq:31.1}. The solution to Eqs.\ \eqref{lambda1} and \eqref{lambda2} gives $\lambda_1$ and $\lambda_2$ in terms of the temperature ratio $\chi_1$. Next, these cumulants are substituted into the steady-state condition ($\Lambda_1^*=0$)
\beq
\label{Lambda1=0}
\frac{2\omega^*\xi^{*1/3}}{M_1^\beta}- \frac{\xi^*}{M_1^{\lambda-1}} +\chi_1 \left(\zeta_{10}+\zeta_{11}\lambda_1+\zeta_{12}\lambda_2\right)=0
\eeq
to get a nonlinear function determining $\chi_1$. This provides entirely all parameters of the scaled distributions $\varphi_i$.

In order to obtain explicit results for the temperature ratio and the cumulants, the form of the pair correlation function $g_{ij}$ must be chosen. Here, as in previous works on granular mixtures, $g_{ij}$  is taken to be the equilibrium pair correlation function. In the case of hard spheres ($d=3$), a good approximation is given by the Carnahan-Starling form \cite{CS}
\beq
\label{gij}
g_{ij}=\frac{1}{1-\phi}+\frac{3}{2}\frac{\eta}{(1-\phi)^2}\frac{\sigma_i \sigma_j}{\sigma_{ij}}   +\frac{1}{2}\frac{\eta^2}{(1-\phi)^3}\left(\frac{\sigma_i \sigma_j}{\sigma_{ij}}\right)^2,
\eeq
where the solid volume fraction $\phi=\phi_1+\phi_2$, and
\begin{equation}
  \label{eq:36}
  \phi_i=\frac{1}{6}n_i\phi \sigma_i^3
\end{equation}
is the species volume fraction of the component $i$. Moreover, in Eq.\ \eqref{gij}, $\eta=\pi(n_1\sigma_1^2+n_2\sigma_2^2)/6$. Comparison with computer simulations for classical binary hard sphere mixtures ($\alpha_{ij}=1$) have shown that the approximation \eqref{gij} turns out to be quite accurate in most of the fluid region, although it fails for high densities and for larger diameter ratios. \cite{SYM02} Given the values considered in our simulations (see below), we expect that Eq.\ \eqref{gij} estimates well the pair correlation function $g_{ij}$ for granular mixtures.

\section{Comparison between theory and molecular dynamics simulations}
\label{sec4}

As said in Sec.\ \ref{sec3}, the approximation \eqref{eq:23} provides detailed predictions for the temperature ratio $\chi_1$ and the cumulants $\lambda_1$ and $\lambda_2$ as functions of the mass ratio $m_1/m_2$, the size ratio $\sigma_1/\sigma_2$, the composition $x_1$, the volume fraction $\phi$, the coefficients of restitution $\alpha_{ij}$, and the driven parameters $\gamma^*$ and $\xi^*$. The degree of reliability of this approximate solution will be assessed in this section via a comparison with MD simulations. This is the main objective of the paper. As already mentioned in the Introduction, in contrast to the DSMC method,\cite{B94} MD simulations avoid any assumptions of kinetic theory (such as, molecular chaos and the explicit form of the Fokker-Planck operator) and hence, the comparison made here can be considered as an stringent test of the domain of validity of the Enskog kinetic theory for conditions of practical interest. Before comparing theory and computer simulations in the steady state, let us first give some technical details of the numerical simulations as well as some comparisons for the time-dependent problem.

\begin{figure}[!h]
\includegraphics[width=.50\columnwidth,angle=-90]{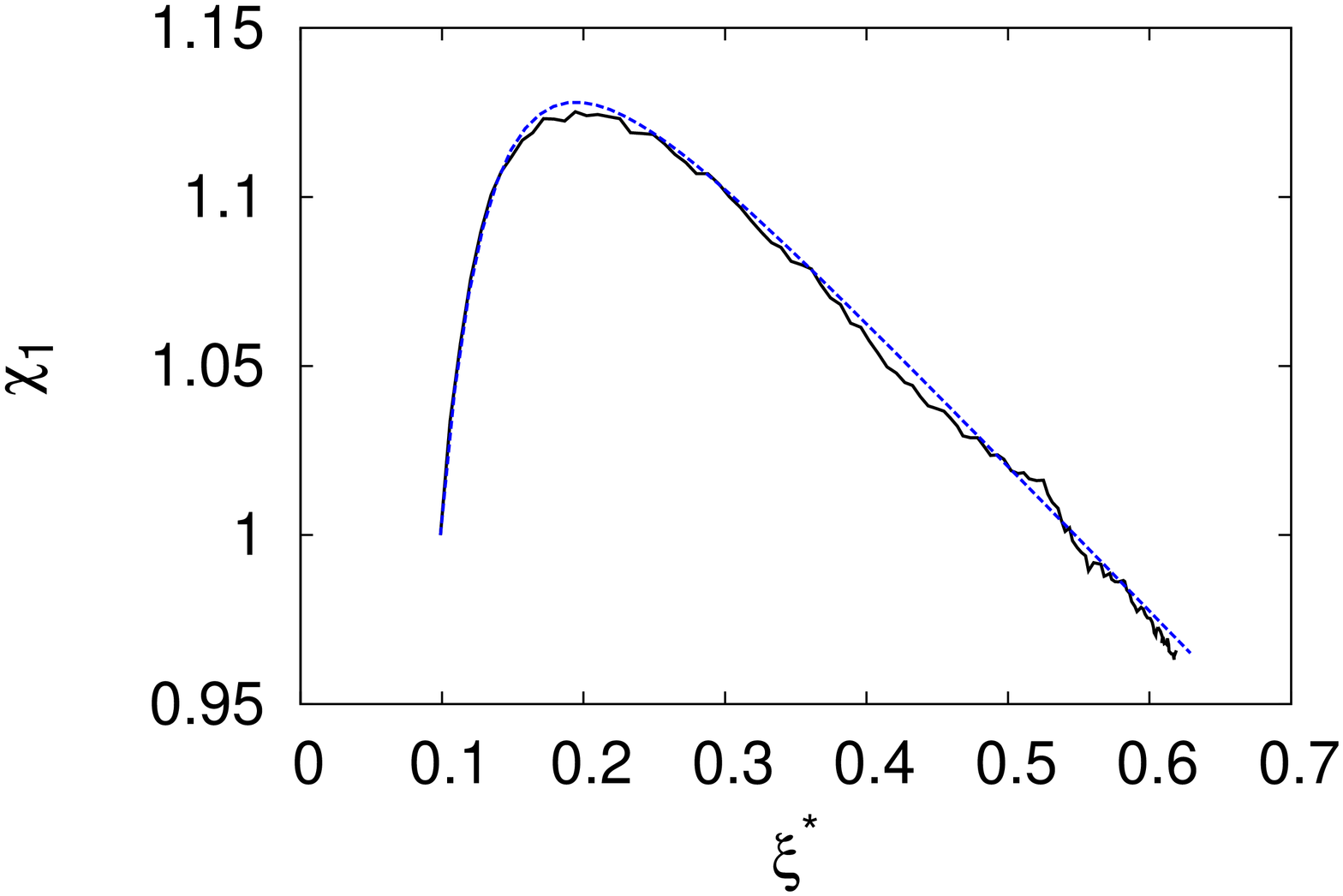}
\includegraphics[width=.48\columnwidth,angle=-90]{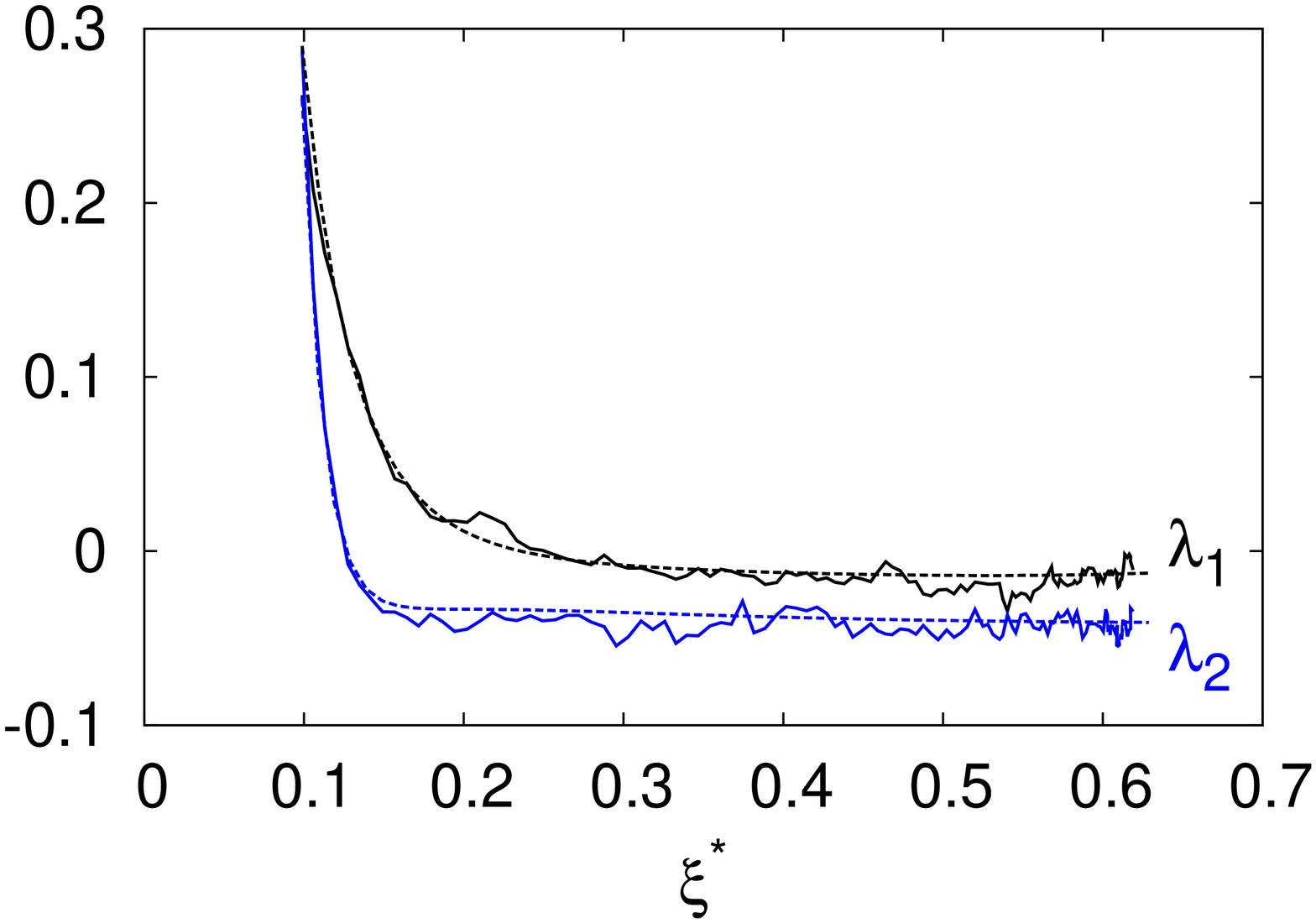}
\caption{(Color online) Dependence of the temperature ratio $\chi_1$ and the cumulants $\lambda_1$ and $\lambda_2$ on the reduced noise strength $\xi^*$ for a volume fraction $\phi=0.00785$ (very dilute system) with $\phi_1=\phi_2=\frac{1}{2}\phi$. Here, $\sigma_1=\sigma_2=0.01\sigma_0$, $m_2=m_0$, $m_1/m_2=10$, and $\alpha_{11}=\alpha_{22}=\alpha_{12}=0.9$. The solid lines correspond to MD simulations while the dashed lines correspond to the theoretical results.}
\label{fig1}
\end{figure}
\begin{figure}[!h]
\includegraphics[width=.50\columnwidth,angle=-90]{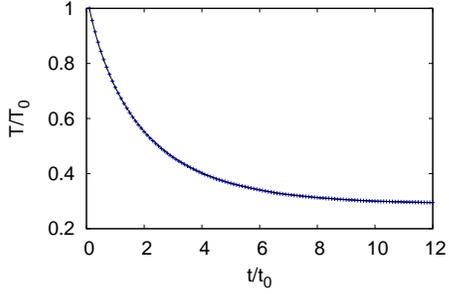}
\caption{(Color online) Time dependence of the (reduced) granular temperature $T/T_0$. The solid line corresponds to the theoretical results while the symbols refer to MD simulations. The parameters of the system are the same as those considered in Fig.\ \ref{fig1}.}
\label{fig2}
\end{figure}

\subsection{Simulation data}

We have simulated via event-driven MD \cite{L91,AT05} a system constituted by a total number of $N=20^3$ inelastic, frictionless hard spheres ($d=3$). The system is inside a box of size $L$ and is subjected to periodic boundary conditions. The granular system under consideration is driven by the action of a deterministic external force proportional to the velocity particle plus a stochastic force. Thus, the velocities of particles of each species change their values between collisions according to the rules \eqref{v1}-\eqref{v2}. In the simulations carried out in this work the parameters of the bath are $\beta=1$, $\lambda=2$, and
\begin{equation}
  \label{eq:35}
\xi_\text{b}^2=0.2 \frac{m_0^2}{\sigma_0}\left(\frac{T_0}{m_0}\right)^{3/2}, \quad \gamma_\text{b}=0.1 \frac{m_0}{\sigma_0}\left(\frac{T_0}{m_0}\right)^{1/2},
\end{equation}
where $m_0$, $\sigma_0$, and $T_0$ are the units of mass, length, and temperature, respectively. The unit of time $t_0$ is $t_0=\sqrt{m_0/(2T_0)}\sigma_0$. The random variable $\textbf{w}$ is uniformly distributed in the interval $(-1,1)$ and $\Delta t$ has been selected in all cases to ensure that it is the smallest time scale in the problem and fulfills inequalities of Sec. \ref{sec2}. In addition, the initial state is the same for almost all simulations, namely (uniform) Gaussian velocity distributions with temperature equal to $T_0$.

\begin{figure}[!h]
\includegraphics[width=.71\columnwidth,angle=-90]{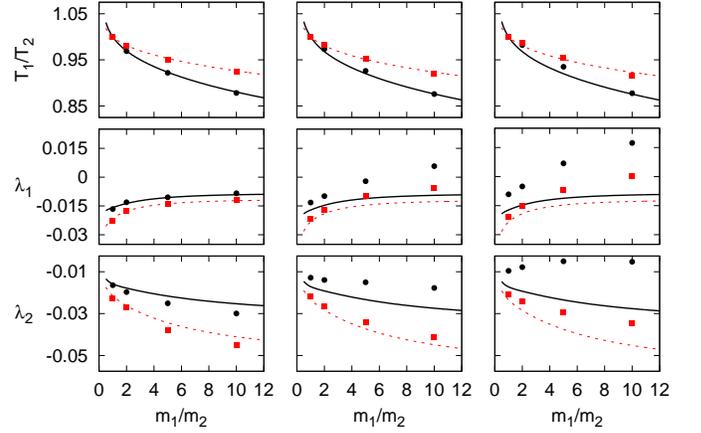}
\caption{(Color online) Case I: Plot of the temperature ratio $T_1/T_2$ and the cumulants $\lambda_1$ and $\lambda_2$ as a function of the mass ratio $m_1/m_2$ for $\sigma_1/\sigma_2=\phi_1/\phi_2=1$, and two different values of the (common) coefficient of restitution $\alpha$: $\alpha=0.8$ (solid lines and circles) and $\alpha=0.9$ (dashed lines and squares). The lines are the Enskog predictions and the symbols refer to the MD simulation results. The first, second and third columns correspond to $\phi=0.00785$, $\phi=0.1$ and $\phi=0.2$, respectively.}
\label{fig3}
\end{figure}
\begin{figure}[!h]
\includegraphics[width=.71\columnwidth,angle=-90]{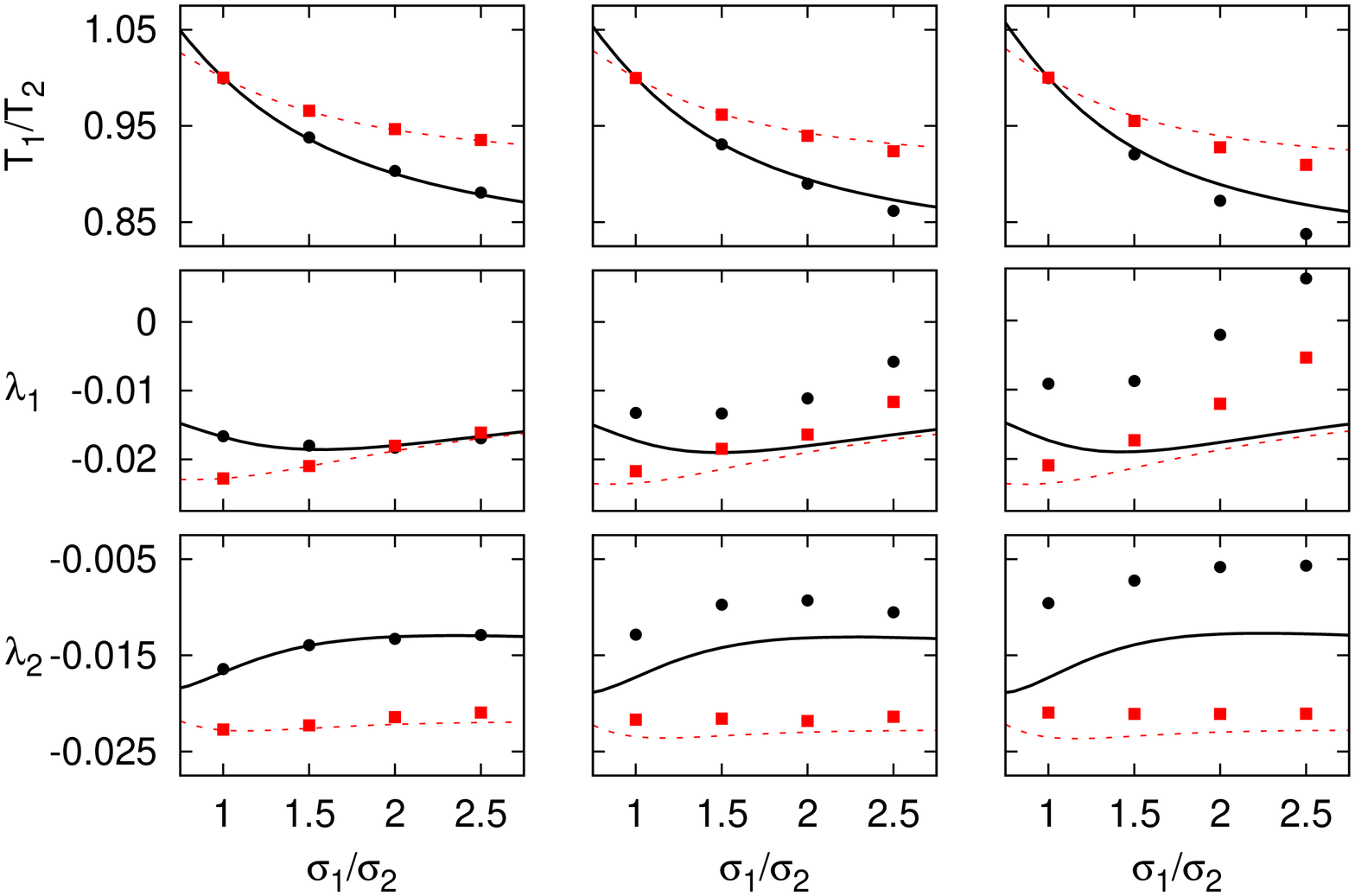}
\caption{(Color online) Case II: Plot of the temperature ratio $T_1/T_2$ and the cumulants $\lambda_1$ and $\lambda_2$ as a function of the size ratio $\sigma_1/\sigma_2$ for $m_1/m_2=\phi_1/\phi_2=1$, and two different values of the (common) coefficient of restitution $\alpha$: $\alpha=0.8$ (solid lines and circles) and $\alpha=0.9$ (dashed lines and squares). The lines are the Enskog predictions and the symbols refer to the MD simulation results. The first, second and third columns correspond to $\phi=0.00785$, $\phi=0.1$ and $\phi=0.2$, respectively.}
\label{fig4}
\end{figure}
\begin{figure}[!h]
\includegraphics[width=.71\columnwidth,angle=-90]{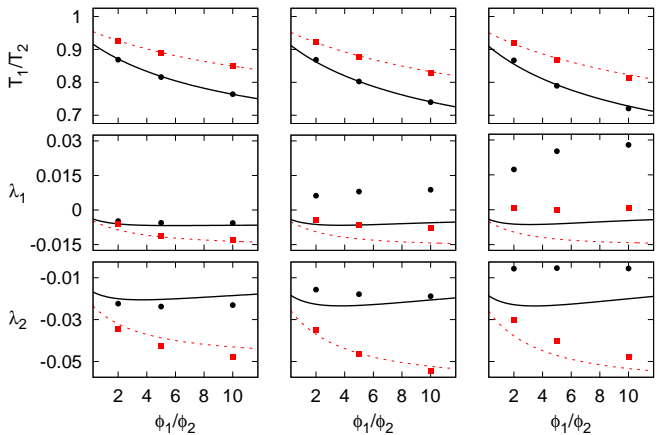}
\caption{(Color online) Case III: Plot of the temperature ratio $T_1/T_2$ and the cumulants $\lambda_1$ and $\lambda_2$ as a function of the composition $\phi_1/\phi_2$ for $m_1/m_2=\sigma_1/\sigma_2=1$, and two different values of the (common) coefficient of restitution $\alpha$: $\alpha=0.8$ (solid lines and circles) and $\alpha=0.9$ (dashed lines and squares). The lines are the Enskog predictions and the symbols refer to the MD simulation results. The first, second and third columns correspond to $\phi=0.00785$, $\phi=0.1$ and $\phi=0.2$, respectively.}
\label{fig5}
\end{figure}

\subsection{Time-dependent states}

Although we are mainly interested in evaluating all the relevant quantities of the problem ($\chi_1$, $\lambda_1$ and $\lambda_2$) in the (asymptotic) steady state, it is also interesting to analyze the approach towards the steady state. Here, for the sake of brevity, we only consider the case of a very dilute mixture ($\phi=0.00785$) with $\phi_1=\phi_2=\frac{1}{2}\phi$. In addition, $\sigma_1=\sigma_2=0.01\sigma_0$, $m_2=m_0$, $m_1/m_2=10$, and $\alpha_{11}=\alpha_{22}=\alpha_{12}=0.9$. Figure \ref{fig1} shows the temperature ratio $\chi_1$ and the cumulants $\lambda_1$ and $\lambda_2$ as functions of the reduced noise strength $\xi^*$ (since $\xi^*\propto T(t)^{-3/2}$, one can take $\xi^*$ instead of $t/t_0$ to analyze the time-dependence of the above quantities). The numerical results have been obtained by averaging over different initial conditions so that, initially, $\xi^*\simeq 0.1$, $\chi_1=1$ and $\lambda_i\simeq 0.3$. The solid lines correspond to the simulations while the dashed lines refer to the numerical solutions of the differential equations \eqref{eq:21} and \eqref{eq:28}. It is quite apparent that the theoretical predictions based on the first Sonine approximation show an excellent agreement with MD simulations. Once $\chi_1$ and the cumulants are known, the time dependence of the granular temperature can be also obtained. This is shown in Fig.\ \ref{fig2} and the (approximate) theory compares very well with computer simulations.

It must be remarked that the results derived in this Subsection for the temperature ratio and the cumulants clearly show that before reaching the steady state the system evolves towards a \emph{universal} hydrodynamic state (independent of the initial conditions) that depends on a new parameter (the reduced noise strength $\xi^*$) measuring the distance to the steady state. As shown in Ref.\ \onlinecite{NG13}, the above unsteady state plays a relevant role in the hydrodynamic description of the system and affects the form of the transport coefficients. This universal character of the (scaled) distribution function has been previously found in some works on \emph{driven} granular gases. \cite{GMT12,AS07}

\subsection{Steady states}

After a transient regime, as expected we observe that the system reaches a steady state for sufficiently long times. In the steady state, the temperature ratio and the cumulants have been calculated over different time registrations ($10^3$ times every $10^2\Delta t$) and over different initial conditions (typically $20$). Moreover, since the parameter space of the problem is large, in order to reduce the number of independent parameters the simplest case of a \emph{common} coefficient of restitution ($\alpha_{11}=\alpha_{22}=\alpha_{12}\equiv \alpha$) is considered. Thus, once the driven parameters are fixed, the parameter space is reduced to five dimensionless quantities: $\left\{m_1/m_2, \sigma_1/\sigma_2, \phi_1/\phi_2, \phi, \alpha \right\}$.

Three different values of the solid volume fractions $\phi$ have been studied here, $\phi=0.00785$, $\phi=0.1$ and $\phi=0.2$. The first system corresponds to a very dilute fluid while the two latter systems represent moderately dense fluids. Two values of the common coefficient of restitution have been considered, $\alpha=0.8$ and $\alpha=0.9$, both representing moderately strong dissipation. The ratio of partial temperatures $T_1/T_2$ and the cumulants $\lambda_1$ and $\lambda_2$ in the \emph{steady} state have been measured for three cases in each state. In the first case (case I), the set of dimensionless parameters $\Xi \equiv \left\{T_1/T_2, \lambda_1, \lambda_2 \right\}$ are obtained as a function of the mass ratio $m_1/m_2$ for $\sigma_1/\sigma_2=\phi_1/\phi_2=1$. The second case (case II) determines $\Xi$ as a function of the diameters ratio $\sigma_1/\sigma_2$ for $m_1/m_2=\phi_1/\phi_2=1$, while the third case (case III) gives $\Xi$ as a function of composition $\phi_1/\phi_2$ for $m_1/m_2=\sigma_1/\sigma_2=1$.

Figure \ref{fig3} shows the results for case I, $\Xi$ as a function of mass ratio. The symbols represent the simulation data where the circles are for $\alpha=0.8$ and the squares are for $\alpha=0.9$. In addition, the plots of the first column for $\Xi$ correspond to $\phi=0.00785$, the plots of the second column correspond to $\phi=0.1$ and the plots of the third column refer to $\phi=0.2$. The Enskog theoretical predictions are given by the solid ($\alpha=0.8$) and dashed ($\alpha=0.9$) lines. The agreement between the theory and the simulation is seen to be very good for the temperature ratio and the cumulants in the low density fluid ($\phi=0.00785$) over the whole range of mass ratios considered and for both values of dissipation. The agreement is also very good for $T_1/T_2$ at moderate densities ($\phi=0.1$ and 0.2), even for extreme values of the mass ratio. This good performance of the Enskog theory for the temperature ratio for moderately driven dense mixtures contrasts with the results obtained in the freely cooling state \cite{DHGD02} where significant discrepancies between the Enskog theory and MD simulations were observed at $\phi=0.2$ (see Fig.\ 2 of Ref. \ \onlinecite{DHGD02}). However, as the second and third columns of Fig.\ \ref{fig3} show, systematic deviations from the Enskog theory for dense mixtures are obtained in the simulations in the case of the cumulants $\lambda_1$ and $\lambda_2$, especially at $\phi=0.2$ for $\lambda_1$.

Figure \ref{fig4} shows the results for case II, $\Xi$ as a function of size ratio. As in Fig.\ \ref{fig3}, the agreement for both $\alpha=0.9$ and $\alpha=0.8$ is excellent in the dilute regime ($\phi=0.00785$), even for the largest size ratio considered. Regarding only the temperature ratio, we see that the Enskog predictions compare very well with simulations, except for the largest size ratio at $\alpha=0.8$ and $\phi=0.2$. For moderate densities, the theoretical values of the cumulants (especially in the case of $\lambda_2$) are smaller than those obtained in the simulations and large differences are observed at $\phi=0.2$. Figure \ref{fig5} shows the results for case III, $\Xi$ as a function of composition. It is quite apparent that the trends are quite similar to those of Figs.\ \ref{fig3} and \ref{fig4}. While good agreement is obtained for the temperature ratio $T_1/T_2$ for all the densities and dissipation, there are significant discrepancies between theory and simulation for $\lambda_1$ and $\lambda_2$ for moderate densities. These differences increase with dissipation (see for instance, the comparison for $\lambda_1$ at $\phi=0.2$ and $\alpha=0.8$).

\section{Discussion}
\label{sec5}

In this paper, granular mixtures in contact with a heat bath have been modeled by the usual inelastic Enskog equation adding a Fokker-Planck term corresponding to a stochastic bath with friction. We have shown that the Fokker-Planck term emerges naturally when the typical frequency collision between grains and bath is big enough. In the case of homogeneous states, the Enskog kinetic equation admits the scaling solution \eqref{eq:12} where the distribution function $\varphi_i$ of each species ($i=1,2$) depends on the granular temperature not only through the (scaled) velocity $\mathbf{c}=\mathbf{v}/v_0(t)$ (as in the HCS \cite{DHGD02}) but also through the (reduced) noise strength $\xi^*$ (defined in Eq.\ \eqref{eq:15}).

On the other hand, in practice, only approximate forms for the distributions $\varphi_i$ are possible and hence, this distribution is represented as an expansion in Sonine polynomials with the leading terms given by Eq.\ \eqref{eq:23}. As in the freely cooling case, \cite{GD99} the weight function (Gaussian) $\varphi_{i,\text{M}}$ for each species is chosen to be scaled relative to the thermal velocity for that species, introducing explicitly the unknown partial temperatures $T_i$. In the steady state, the ratio of partial temperatures $T_1/T_2$ and the cumulants $\lambda_i$ have been explicitly determined as functions of the mass and size ratios, the composition, the volume fraction and the coefficients of restitution.

The theoretical predictions for $T_1/T_2$ and $\lambda_i$ have been tested against MD simulations for conditions covering dilute and moderate densities as well as moderate and strong dissipation. As Figs.\ \ref{fig3}-\ref{fig5} clearly show, the results of the Enskog equation for the temperature ratio agree very well with MD results for all the systems considered in the simulations. This good agreement contrasts with the comparison carried out in the HCS \cite{DHGD02} where significant discrepancies for $T_1/T_2$ were observed for moderate densities. With respect to the cumulants $\lambda_i$, the theory compares quite well in the low-density regime but systematic significant deviations appear as the density increases. It is important to note that although the evaluation of the temperature ratio involves the knowledge of cumulants (see Eq.\ \eqref{Lambda1=0}), the latter quantities are in general very small and hence, they can be neglected in the evaluation of $T_1/T_2$. In this sense, while the test of the Enskog equation for the temperature ratio is actually an assessment of the Enskog predictions of the cooling rates (which are essentially transport properties and so, they appear in the hydrodynamic equations), the test of the cumulants (which are related with the fourth-degree velocity moments of the scaled distributions $\varphi_i$) is a more stringent comparison than the partial temperatures since they provide an indirect information on the high velocity population of the distributions $\varphi_i$.

As already mentioned in previous works, \cite{DHGD02} the failure of the Enskog theory at high densities for the cumulants can be expected from experience with ordinary (elastic) fluids. This is due to multiparticle collisions that may be stronger for fluids with inelastic collisions where the colliding pairs tend to be more focused. In this context, it is possible that the range of densities for which the Enskog kinetic theory holds decreases with increasing dissipation. This has been already observed \cite{enskog} in some previous comparisons. However, despite this limitation, the Enskog equation can be still considered as a remarkable equation for describing macroscopic properties (such as transport coefficients) for fluids with elastic and inelastic collisions, including mixtures. Recent results for instabilities of granular flows at moderate densities in monocomponent \cite{peter} and binary mixtures \cite{MGH14} confirm the above expectation.

\acknowledgments
The present work has been supported by the Spanish Government through grant No. FIS2010-16587, partially financed by
FEDER funds and by the Junta de Extremadura (Spain) through Grant No. GRU10158.

\appendix
\section{Expressions of $\zeta_{ij}$ and $\Sigma_{ij}$}
\label{appen1}

In this Appendix, the expressions of the cooling rates $\zeta_i^*$ and the fourth degree collisional moments $\Sigma_i$ for a $d$ dimensional granular mixture are given. The procedure to determine them is quite similar to the one previously carried out for hard spheres in Ref.\ \onlinecite{GD99}. Here, for the sake of brevity, we only display the final results.

By using the leading Sonine approximation \eqref{eq:23} for $\varphi_1$ and neglecting nonlinear terms in $\lambda_1$ and $\lambda_2$, the (reduced) partial cooling rate $\zeta_1^*$ can be written in the form of Eq.\ \eqref{eq:31} where \cite{GFM09}
\begin{eqnarray}
\label{a11}
\zeta_{10}&=&\frac{\sqrt{2}\pi^{(d-1)/2}}{d\Gamma\left(\frac{d}{2}\right)}x_1 g_{11}\left(\frac{\sigma
_1}{\sigma_{12}}\right)^{d-1}\theta_1^{-1/2}
(1-\alpha_{11}^2)\nonumber\\
& & +\frac{4\pi^{(d-1)/2}}{d\Gamma\left(\frac{d}{2}\right)}x_2 g_{12}\mu_{21} \left(\frac{1+\theta}{\theta}\right)^{1/2}(1+\alpha_{12})\theta_2^{-1/2}
\nonumber\\
& & \times
\left[1-\frac{1}{2}\mu_{21}(1+\alpha_{12})(1+\theta) \right],
\end{eqnarray}
\begin{eqnarray}
\label{a12}
\zeta_{11}&=&\frac{3\pi^{(d-1)/2}}{16\sqrt{2}d\Gamma\left(\frac{d}{2}\right)} x_1 g_{11} \left(\frac{\sigma
_1}{\sigma_{12}}\right)^{d-1} \theta_1^{-1/2}
(1-\alpha_{11}^2)\nonumber\\
& & +\frac{\pi^{(d-1)/2}}{8d\Gamma\left(\frac{d}{2}\right)}x_2 g_{12}\mu_{21}
\frac{(1+\theta)^{-3/2}}{\theta^{1/2}}(1+\alpha_{12})
\theta_2^{-1/2} \nonumber\\
& & \times
\left[2(3+4\theta)-3\mu_{21}(1+\alpha_{12})(1+\theta) \right],
\end{eqnarray}
\begin{eqnarray}
\label{a13} \zeta_{12}&=&-\frac{\pi^{(d-1)/2}}{8d\Gamma\left(\frac{d}{2}\right)}x_2 g_{12}
\mu_{21}\left(\frac{1+\theta}{\theta}\right)^{-3/2}(1+\alpha_{12})\theta_2^{-1/2}
\nonumber\\
& & \times
\left[2+3\mu_{21}(1+\alpha_{12})(1+\theta_{12}) \right].
\end{eqnarray}
Here, $\theta=\theta_1/\theta_2=m_1 T_2/m_2 T_1$. The partial cooling rate $\zeta_2^*=\zeta_{20}+\zeta_{22}\lambda_2+\zeta_{21} \lambda_1$, where the forms of $\zeta_{20}$, $\zeta_{22}$ and $\zeta_{21}$ can be easily inferred from Eqs.\ (\ref{a11})--(\ref{a13}) by interchanging 1 and 2 and setting $\theta\to \theta^{-1}$.

The fourth degree collisional moment $\Sigma_1$ can be written in the form \eqref{eq:31.1} where \cite{GFM09}
\begin{eqnarray}
\label{a17}
\Sigma _{10} &=&-\frac{\pi^{(d-1)/2}}{\sqrt{2}\Gamma\left(\frac{d}{2}\right)} \theta_{1}^{-5/2}
 x_{1}g_{11}\left(\frac{\sigma_{1}}{{\sigma}_{12}}\right)^{d-1} \frac{ 3+2d+2\alpha_{11}^{2}}{2}
\nonumber\\
& & \times \left(1-\alpha_{11}^{2}\right)+\frac{\pi^{(d-1)/2}}{\Gamma\left(\frac{d}{2}\right)} \theta_{1}^{-5/2}x_2 g_{12}\left( 1+\theta\right)^{-1/2}\nonumber\\
& & \times  \mu_{21}\left( 1+\alpha_{12}\right)\left\{ -2\left[d+3+(d+2)\theta\right] +\mu_{21}\right.\nonumber\\
& & \times \left( 1+\alpha_{12}\right) \left( 1+\theta \right)
\left( 11+ d+\frac{d^2+5d+6}{d+3} \theta \right)\nonumber\\
& & -8\mu _{21}^{2}\left( 1+\alpha_{12}\right)^{2}\left( 1+\theta \right)^{2} +2\mu_{21}^{3}\left(
1+\alpha_{12}\right) ^{3}\nonumber\\
& & \left. \times \left( 1+\theta \right)^{3}\right\},
\end{eqnarray}
\begin{eqnarray}
\label{a18}
\Sigma _{11} &=&-\frac{\pi^{(d-1)/2}}{\sqrt{2}\Gamma\left(\frac{d}{2}\right)} \theta_{1}^{-5/2}
 x_{1}g_{11} \left(\frac{\sigma_{1}}{{\sigma}_{12}}\right)^{d-1} \left[\frac{d-1}{2}(1+\alpha_{11})\right.\nonumber\\
 & & \left.+ \frac{3}{64}
\left( 10d+39+10\alpha_{11}^{2}\right) \left( 1-\alpha_{11}^{2}\right)\right] \nonumber \\
& & +\frac{\pi^{(d-1)/2}}{16\Gamma\left(\frac{d}{2}\right)} \theta_{1}^{-5/2}x_2 g_{12}\left(1+\theta\right)^{-5/2}
\mu_{21}\left(1+\alpha_{12}\right)\nonumber\\
& & \times\left\{-2\left[ 45+15d+(114+39d)\theta
+(88+32d)\theta^{2}\right.\right.\nonumber\\
& & \left.+(16+8d)\theta^{3}\right]
+3\mu_{21}\left( 1+\alpha_{12}\right) \left( 1+\theta \right) \left[ 55+5d \right.\nonumber\\
& & \left. +9(10+d)\theta+4(8+d)\theta
^{2}\right]-24\mu_{21}^{2}\left( 1+\alpha_{12}\right)^{2}
\nonumber\\
& & \left.\times
\left( 1+\theta \right) ^{2}\left( 5+4\theta
\right)+30\mu_{21}^{3}\left( 1+\alpha_{12}\right)^{3}\left( 1+\theta \right)^{3}\right\},\nonumber\\
\end{eqnarray}
\begin{eqnarray}
\label{a19}
\Sigma_{12} &=&\frac{\pi^{(d-1)/2}}{16\Gamma\left(\frac{d}{2}\right)}
\theta_{1}^{-5/2}x_2 g_{12}\theta^2\left(1+\theta_{12}\right)^{-5/2} \mu_{21}\left(
1+\alpha_{12}\right) \nonumber\\
& & \times
\left\{ 2\left[ d-1+(d+2)\theta_{12} \right] +3\mu _{21}\left( 1+\alpha
_{12}\right) \left( 1+\theta\right)\right.
\nonumber\\
& & \left.\times \left[d-1+(d+2)\theta\right]
-24\mu _{21}^{2}\left( 1+\alpha_{12}\right)^{2}\left( 1+\theta \right) ^{2}\right.
\nonumber\\
& & \left.
+30\mu_{21}^{3}\left(1+\alpha_{12}\right) ^{3}\left( 1+\theta \right) ^{3}\right\}  \;.
\end{eqnarray}
The fourth degree collisional moment $\Sigma_2=\Sigma_{20}+\Sigma_{22}\lambda_2+\Sigma_{21}\lambda_1$, where as before
the expressions for $\Sigma_{20}$, $\Sigma_{22}$ and $\Sigma_{21}$ are easily obtained from Eqs.\ (\ref{a17})--(\ref{a19}) by changing $1 \to 2$ and $\theta\to \theta^{-1}$.

In the case of mechanically equivalent particles ($\sigma_1=\sigma_2$, $m_1=m_2$, $\alpha_{11}=\alpha_{22}=\alpha_{12}$), Eqs.\
(\ref{a11})--(\ref{a13}) and \eqref{a17}--\eqref{a19} are consistent with those previously obtained for a single gas. \cite{NE98} Also, for $d=3$, the expressions (\ref{a11})--(\ref{a13}) and \eqref{a17}--\eqref{a19} agree with those derived for a binary mixture of inelastic hard spheres. \cite{GD99} This shows the consistency of the general expressions displayed here.

\end{document}